\documentclass[english,12pt]{article}
\usepackage{color}
\usepackage{amssymb}
\usepackage{amsmath}
\usepackage{babel}
\usepackage{pifont}
\usepackage
[dvips,
colorlinks=true,
linkcolor=webgreen,
filecolor=webbrown,
citecolor=webgreen,
bookmarksopen=false,
]{hyperref}
\definecolor{webgreen}{rgb}{0,.5,0}
\definecolor{webbrown}{rgb}{.6,0,0}
\usepackage[T1]{fontenc}
\usepackage[cp1250]{inputenc}
\usepackage{array}
\usepackage[dvips]{graphicx}
\usepackage{amssymb}
\date{}
\definecolor{arcolor}{cmyk}{0.05,0.95,0.9,0.1}
\title{Interference of Quantum Market Strategies}
\author{Edward W. Piotrowski\\ Institute of Theoretical Physics,
University of Bia\l ystok,\\ Lipowa 41, Pl 15424 Bia\l ystok,
Poland\\ e-mail: \href{mailto:ep@alpha.uwb.edu.pl}{ep@alpha.uwb.edu.pl}\\
 Jan S\l adkowski \\ Institute of Physics, University of Silesia, \\ Uniwersytecka
4, Pl 40007 Katowice, Poland \\ e-mail:
\href{mailto:sladk@us.edu.pl}{sladk@us.edu.pl} \\ and Jacek Syska\\
Institute of Physics, University of Silesia, \\ Uniwersytecka 4,
Pl 40007 Katowice, Poland \\ e-mail:
\href{mailto:jacek@server.phys.us.edu.pl}{jacek@server.phys.us.edu.pl}}
\begin{document}
\maketitle
\def\Z{{\bf Z\!\!Z}}
\def\R{{\bf I\!R}}
\def\N{{\bf I\!N}}
\begin{abstract}
Recent development in quantum computation and quantum information
theory  allows to extend the scope of game theory for the quantum
world. The paper is devoted to the analysis of interference of
quantum strategies in quantum market games.

\end{abstract}


 \vspace{5mm}

\section{Introduction}   Recent development in quantum computation
and quantum information theory allows to extend the scope of game
theory for the quantum world \cite{1,2}. In the "standard" quantum
game theory one tries in some sense to quantize an operational
description of the "classical" game in question. Iqbal and Toor
have successfully applied the method of quantization of games
proposed by Marinatto and Weber \cite{3} in biology \cite{4,5,6}
and recently they have used the same formalism to analyse the
Stackelberg duopoly (leader-follower model) \cite{7}. In the
classical setting of the duopoly game the follower becomes
worse-off compared to the leader who becomes better-off. But they
have shown that in the quantum version the follower is not hurt
even if he or she knows the action of the leader. The backward
induction outcome is the same as the Nash equilibrium in the
classical Cournot game (that is when decisions are made
simultaneously and there is no information hurting players). Two
of the present authors have managed to formulate a new approach to
quantum game theory that is suitable for description of market
transactions in terms of supply and demand curves
\cite{8}-\cite{11}. In this approach quantum strategies are
vectors in some Hilbert space and can be interpreted as
superpositions of trading decisions. For a trader they form the
"quantum board". The actual subject of investigation may consist
of single traders, teams of traders or even the whole market. Not
the apparatus nor installation for actual playing but strategies
are at the very core of the theory. Spontaneous or
institutionalized market transactions are described in terms of
projective operations acting on Hilbert spaces of traders'
strategies. Quantum entanglement is necessary to strike the
balance of trade. The discussed below text-book examples of
departures from the demand and supply law are related to the
negative probabilities that often emerge in quantum theories and
form very interesting illustrations of them \cite{9,10}. This
model predicts the property of undividity of the traders'
attention (no clonning theorem).
The theory unifies also the English auction with the Vickrey's one
attenuating the motivation properties of the latter. There are
analogies with quantum thermodynamics that allow to interpret
market equilibrium as a state with vanishing financial risk flow.
Sometimes euphoria, panic or the herd instinct cause violent
changes of market prices. Such phenomena can be described by the
non-commutative quantum mechanics \cite{8,10}. There is a simple
tactic that maximize the trader's profit on an effective market:
{\it accept profits equal or greater then the one you have
formerly achieved on average}. Here we would like to analyse the
consequences of interference of quantum strategies. We begin with
the analysis of tactics changing the demand. Then we discuss the
supply-side aspects. Both sides' strategies are related by the
Fourier transform.

\section{Tactics changing demand} Traders' strategies are
represented by probability amplitudes (wave functions)
$|\psi\rangle\negthinspace$. There are two important
representations $\langle q|\psi\rangle\negthinspace$ (demand
representation) and $\langle p|\psi\rangle\negthinspace$ (supply
representation) where $q$ and $p$ are logarithms of prices at
which the player is buying or selling, respectively
\cite{8,11,12}.  We have defined canonically conjugate hermitian
operators  of demand $\mathcal{Q}$ and supply $\mathcal{P}$
corresponding to the variables $q$ and $p$ characterizing strategy
of  the  player. This led us to the definition of the observable
that we call {\it the risk inclination operator} \cite{8}:
$$
H(\mathcal{P},\mathcal{Q}):=\frac{(\mathcal{P}-p_{0})^2}{2\,m}+
                     \frac{m\,\omega^2(\mathcal{Q}-q_{0})^2}{2}\,,
\eqno(2) $$ \noindent where $p_{0}:=\frac{
\phantom{}\negthinspace\langle\psi|\mathcal{P}|\psi\rangle}
{\phantom{}\negthinspace\langle\psi|\psi\rangle}\,$,
$q_{0}:=\frac{
\phantom{}\negthinspace\langle\psi|\mathcal{Q}|\psi\rangle }
{\phantom{}\negthinspace\langle\psi|\psi\rangle}\,$,
$\omega:=\frac{2\pi}{\theta}\,$.  $ \theta$  is, roughly speaking,
an average time spread between two opposite moves of a player
(e.g.~buying and selling the same commodity) \cite{12}. The
parameter $m\negthinspace>\negthinspace0$ measures the risk
asymmetry between buying and selling positions. Analogies with
quantum harmonic oscillator are not accidental and one can even
define a parameter $h_E$ that describes the minimal inclination of
the player to risk \cite{8}. The distribution functions, that is
integrals of squared absolute values of the amplitudes, have
natural interpretation in terms of demand and supply curves
\cite{10,11}. This picture should be supplemented with dynamics
resulting from the traders' tactical moves. Tactical moves of
traders, that modify their behaviour on the market, are described
by unitary operations on the Hilbert space of strategies $ L^2$,
$\langle q|\psi\rangle\negthinspace\in\negthinspace L^2$. The
fundamental tactic that switches the supply strategy into the
demand one and vice versa is described by the Fourier transform.
There are also tactics that only modify the demand or supply
strategy without switching sides. We are interested in such
tactics that the resulting strategy is a pure state and therefore
cannot be reduced to a convex combination of strategies that is to
the components having definite supply or demand characteristics.
Such combinations are possible only if components are not
orthogonal (that is, they do not correspond to classical
strategies). The possibility of forming such combinations is a
pure quantum effect that cannot be described in terms of the
classical probabilistic
approach. \\
The "linear shift" of the demand by $\Delta$
\begin{equation*}
\mathcal{D}_\Delta\langle q|\psi\rangle:=\langle
q-\Delta|\psi\rangle\,
\end{equation*}
commutes with the supply operator $\mathcal{P}$ \cite{8,10} and
therefore does not change the supply side. It only results in a
phase shift of the Fourier transformed strategy, $\langle
p|\psi\rangle\rightarrow {\mathrm
e}^{\mathrm{i}\hslash_{E}^{-1}\Delta\,p}\, \langle p|\psi\rangle$,
so the shape of the supply curve of the player is unchanged. The
quantum formalism enables playing by using families of tactics of
the form (the proper normalization can be achieved by respective
normalization of supply and demand curves)
\begin{equation*}
\mathcal{D}_\Delta(\xi_0,\xi_1):=\xi_0
\mathcal{I}+\xi_1\mathcal{D}_\Delta ,
\end{equation*}
where $(\xi_0,\xi_1)$ are homogeneous complex projective
coordinates of points on the Riemann sphere
$S^2\simeq\mathbb{C}P^1\simeq\overline{\mathbb{C}}$, and
$\mathcal{I}=\mathcal{D}_0$ is the identity map. The poles $(1,0)$
and $(0,1)$ correspond to identity mapping of the initial strategy
and its shifted copies, respectively. The remaining points
parameterize all possible superpositions of the initial
strategies. We define the (commuting) composition of tactics
transforming the initial strategy $\langle q|\psi\rangle$ as:
\begin{equation*}
{\mathcal{D}_{\Delta'}}(\xi_0',\xi_1')\circ{\mathcal{D}_{\Delta''}}(\xi_0''
,\xi_1''):=
\mathcal{D}_{\Delta'+\Delta''}(\xi_0'\cdot\xi_0'',\xi_1'\cdot\xi_1'')\,.
\end{equation*}
The transition from a state being a normal (Gaussian) distribution
to a superposition of such states is not unitary but the tactics
$\mathcal{D}_\Delta(\xi_0\neq0,\xi_1\neq0)$ manipulating
coefficients of the superposition or the widths $\Delta$ of
Gaussian components are unitary so it is convenient to use instead
of the tactics $\mathcal{D}(1,0)$ and $\mathcal{D}_\Delta(1,0)$
series of tactics which are convergent to them.

Let us investigate the effect of the family of tactics
${{\mathcal{D}_{\Delta}}(\xi_0,\xi_1)}_{\xi_0,\xi_1}$ on the
process of changing the demand (and the corresponding supply) with
the effects of the interference of the strategies. They represent
behaviours of the player which differ from the standard one
dictated by the law of demand and supply. The diverse market
strategies might be identified with the behaviours  of the player
connected with the price of the particular goods (which have
appeared under different states of his (or her) knowledge or
perhaps subconsciousness). Then the superposition of the
strategies of the player could be interpreted as the appearance of
the diverse types of couplings between simultaneous ensembles of
different states of the player. Often during the alteration of our
opinions and characteristic behaviours, the old and new stages of
our changing individuality coexist with each other in a variety of
relations. With the lack of such coexistence, "schizophrenic"
states might lead to the behaviors which are socially inefficient.

\section{The results of the tactic
${\mathcal{D}_{\Delta}}(\xi_0,\xi_1)$ for the Gaussian strategy}

We are interested in quantum description of a game which consists
in buying or selling of market goods. We will investigate the pure
adiabatic strategies only. Such a restriction is justified by the
fact that in the market reality the inclination to the risk seems
to be a characteristic feature of the player hence for one
participant of the market it is unchanged even over big intervals
of the time. Therefore the player has at his (her) disposal the
Gaussian standard strategy, which is equal to the square root of
the standard normal distribution $\langle q|\psi\rangle_{g} :=
\frac{1}{\sqrt[4]{2 \pi }}\, {\mathrm e}^{-\frac{q^2}{4}}$ as a
function of the logarithm of the price, above which the player
give up buying goods offered to him (withdrawal price
\cite{10,12}). The free choice both of the currency unit and the
logarithmic base allows to choose the two first moments of the
distribution $|\langle q|\psi\rangle_g|^2$ to be equal to
$q_0\negthinspace=\negthinspace0$ and
$\sigma\negthinspace=\negthinspace1$ respectively. It is worth to
notice that the Gaussian strategy is, in the quantum setting, the
only one which fulfills the law of the demand and supply.

Let the player operates with  tactics which belonging to the
family
$\{\mathcal{D}_\Delta(z):=\mathcal{D}_\Delta(\xi_0,\xi_1)\}$. To
abbreviate the notation we use the inhomogeneous coordinate on the
Riemann sphere, $\frac{\xi_1}{\xi_0}\negthinspace=:\negthinspace
z\negthinspace\in\negthinspace \overline{\mathbb{C}}$. As the
result of applying the tactics $\mathcal{D}_\Delta(z)$, the
Gaussian strategy is replaced with the superposition of the
Gaussian strategies, given by
\begin{equation*}
\mathrm{e}^{-\frac{q^2}{4}}\,\longrightarrow\,
\mathrm{e}^{-\frac{q^2}{4}}+ z\,
\mathrm{e}^{-\frac{(q-\Delta)^2}{4}} \; .
\end{equation*}

\begin{figure}[h]
\begin{center}
\includegraphics[height=6.25cm, width=9cm]{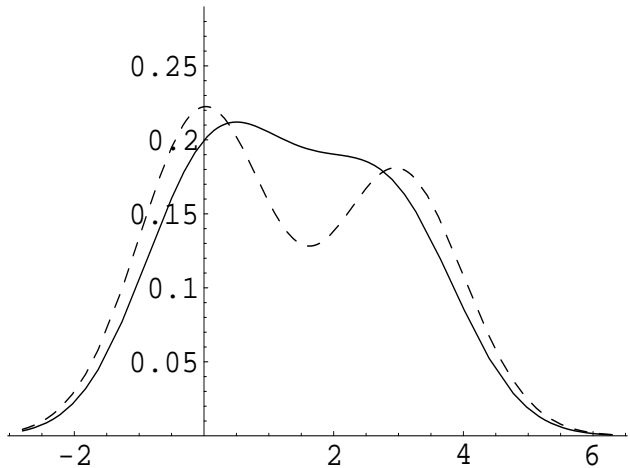}
\end{center}
\caption{Constructive interference of two Gaussian strategies
which are the effect of the tactic $\mathcal{D}_3(0.9)$. The
dashed line corresponds to the tactic
$\mathcal{D}_3(\pm0.9\text{i})$.} \label{interjed}
\end{figure}

In the following Figures we will compare the player's demand given
by the function of the probability density distribution
$|\mathcal{D}_\Delta(z)\langle q|\psi\rangle_g|^2$ with the
density function given by\footnote{In the quantum theory such a
type of  density functions appear  in the case of the tactics
$\mathcal{D}_\Delta(z)$, for which $z$ is  imaginary
($\arg(z)=\pm\tfrac{\pi}{2}$). }
\begin{equation}
\label{gesttosc}
\frac{1}{\sqrt{2\pi}}\Bigl(\frac{1}{1+|z|^2}\,\mathrm{e}^{-\frac{q^2}{2}}+
\frac{|z|^2}{1+|z|^2}\,\mathrm{e}^{-\frac{(q-\Delta)^2}{2}}\Bigr)\,,
\end{equation}
which would appear in the classical theory (in which the
interference effects are neglected). As a result of interference,
for $\arg(z)\negthinspace=\negthinspace0$, we observe the biggest
overlapping of two Gaussian curves, which in addition are
submitted to a mutual attraction. Figure~\ref{interjed}
illustrates the case for $\Delta\negthinspace=\negthinspace3$ and
$z\negthinspace=\negthinspace0.9$. In order to demonstrate the
scale of the quantum effects, the corresponding sum of two
Gaussian strategies which does not have the interference
ingredient $(\ref{gesttosc})$ is drawn in Figure~\ref{interjed}
and the following ones as the dashed line.
\begin{figure}[h]
\begin{center}
\includegraphics[height=6.25cm, width=9cm]{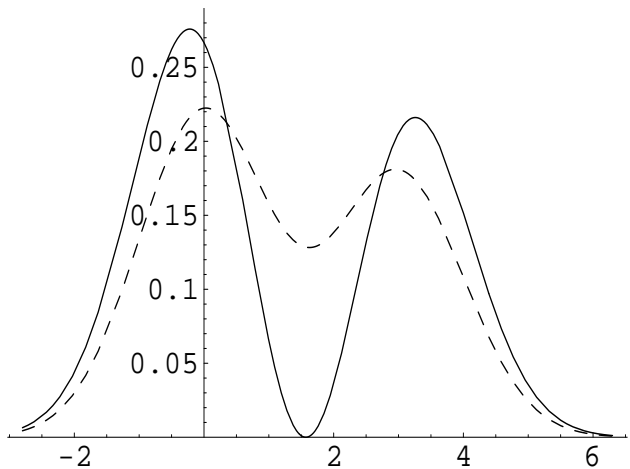}
\end{center}
\caption{Destructive interference of two Gaussian strategies for
the tactic $\mathcal{D}_3(-0.9)$. The dashed line corresponds to
the tactic $\mathcal{D}_3(\pm0.9\text{i})$.} \label{interdwa}
\end{figure}

\begin{figure}[h]
\begin{center}
\includegraphics[height=6.25cm, width=9cm]{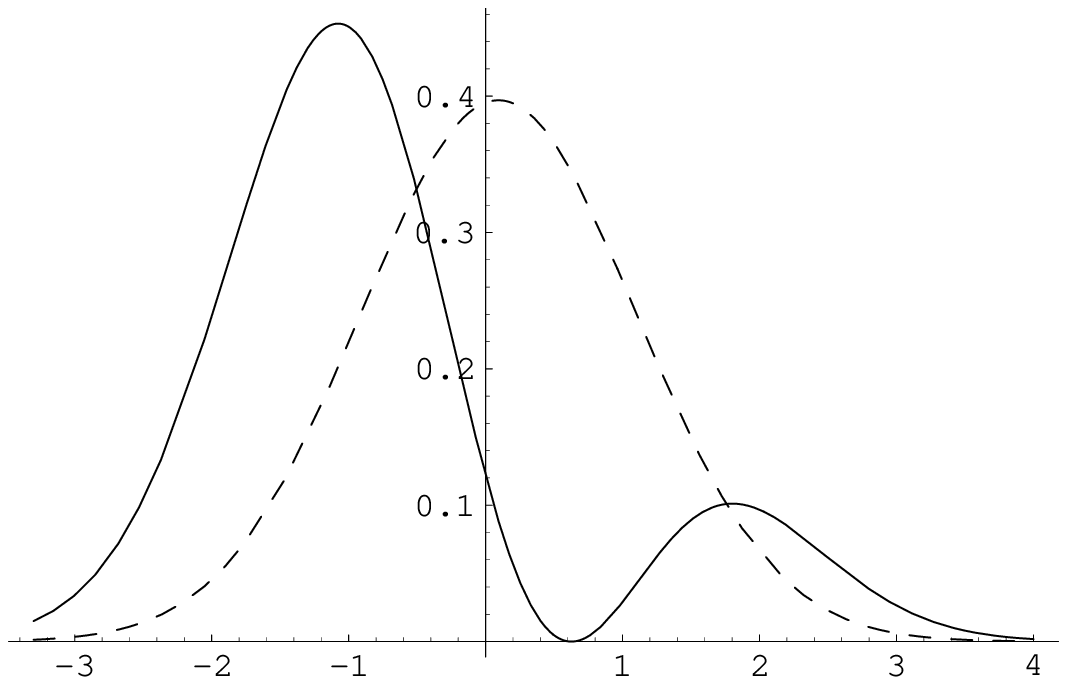}
\end{center}
\caption{The destructive interference "pushing aside" two of the
overlapping Gaussian strategies having similar height -- the
tactic $\mathcal{D}_{0.2}(-\sqrt{0.9})$. The dashed line
corresponds to the tactic
$\mathcal{D}_{0.2}(\pm\sqrt{0.9}\text{i})$.} \label{tertrz}
\end{figure}

In contrast is the case with
$\arg(z)\negthinspace=\negthinspace\pi$. It is represented in
Figure~\ref{interdwa}. Now, after the change of the phase of the
$z$ parameter only, the bell like curves which have their origin
in the superposition of the Gaussian curves are fully separated.
The interference is also the cause of their mutual repulsion. The
smaller  the gap between the Gaussian components the stronger the
influence of the interference is.  Figure~\ref{tertrz} represents
the case with the Gaussian curves of similar height
($z\negthinspace=\negthinspace-\sqrt{0.9}$) which are overlapping
in practice ($\Delta\negthinspace=\negthinspace0.2$). Yet, the
destructive interference enhances these features that discriminate
between the curves.

\begin{figure}[h]
\begin{center}
\includegraphics[height=6.25cm, width=9cm]{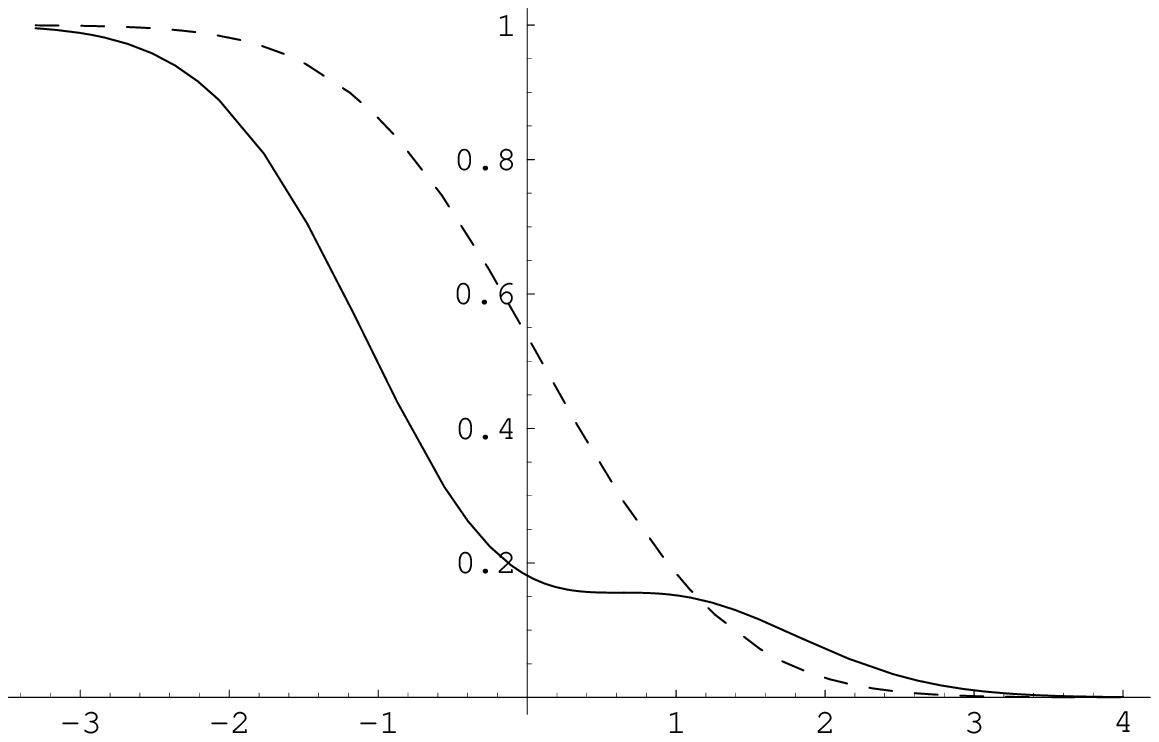}
\end{center}
\caption{The change of the demand curve caused by the quantum
interference effect presented in  Figure~\ref{tertrz}.}
\label{interczt}
\end{figure}

The integral curves of the distribution of the probability
densities given in Figure~\ref{tertrz} (hence the cumulative
probability, but with the inverted domain) are represented in
Figure~\ref{interczt}. In  agreement with the previously given
market interpretation of the quantum theory \cite{9}-\cite{11}
they describe the influence of the quantum (interference) effects
onto the shape of the demand curve calculated for the tactic
$\mathcal{D}_{0.2}(-\sqrt{0.9})$.

\section{Fourier transform of the tactics
$\mathcal{D}_{\Delta}(z)$}

We have analyzed the demand  following from the strategy which in
the demand representation is a superposition of the Gaussian
strategies. The description of the strategies in terms of squared
modules of the probability amplitudes (as it is in the classical
picture) and not the amplitudes themselves is incomplete. Hence it
is interesting to complete it by the description of the very
strategy in the relation to its supply aspect. The quantum
description of the market presented in this paper requires to
perceive a player as a subject of a market, who according to the
demand representation $\langle q|\psi \rangle$ of his strategy
$|\psi\rangle$ acquires a specific commodity and consuming it
(processing or leaving unchanged) gains a profit described by the
operator $\mathcal{Q}+\mathcal{P}$. It is this observable, and not
$\mathcal{Q}$ alone, which does not depend on the choice of the
monetary unit and measures the price of the commodity in the
transaction. The density   of the probability of the supply random
variable $\mathfrak{p}$ determines the square of the module of the
supply representation $\langle p|\psi \rangle$ of the same
strategy $|\psi\rangle$ according to which the goods has been
purchased!

\begin{figure}[h]
\begin{center}
\includegraphics[height=6.25cm, width=9cm]{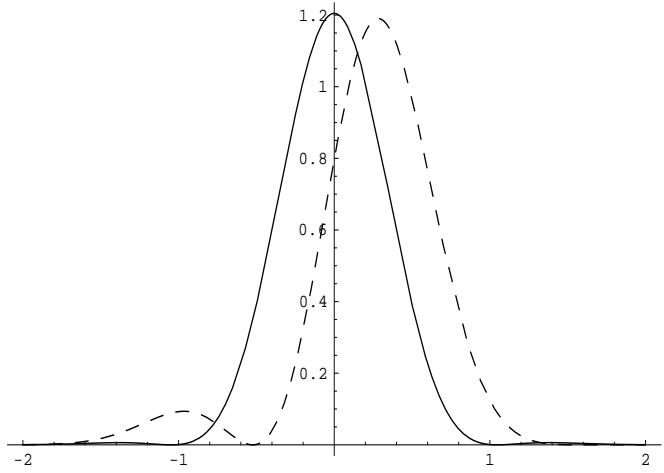}
\end{center}
\caption{The probability density for the Fourier transforms  of
the tactics $\mathcal{D}_{3}(1)$ (the continuous line) and
$\mathcal{D}_{3}(-\text{i})$ (the dashed line).} \label{intosi}
\end{figure}

\begin{figure}[h]
\begin{center}
\includegraphics[height=6.25cm, width=9cm]{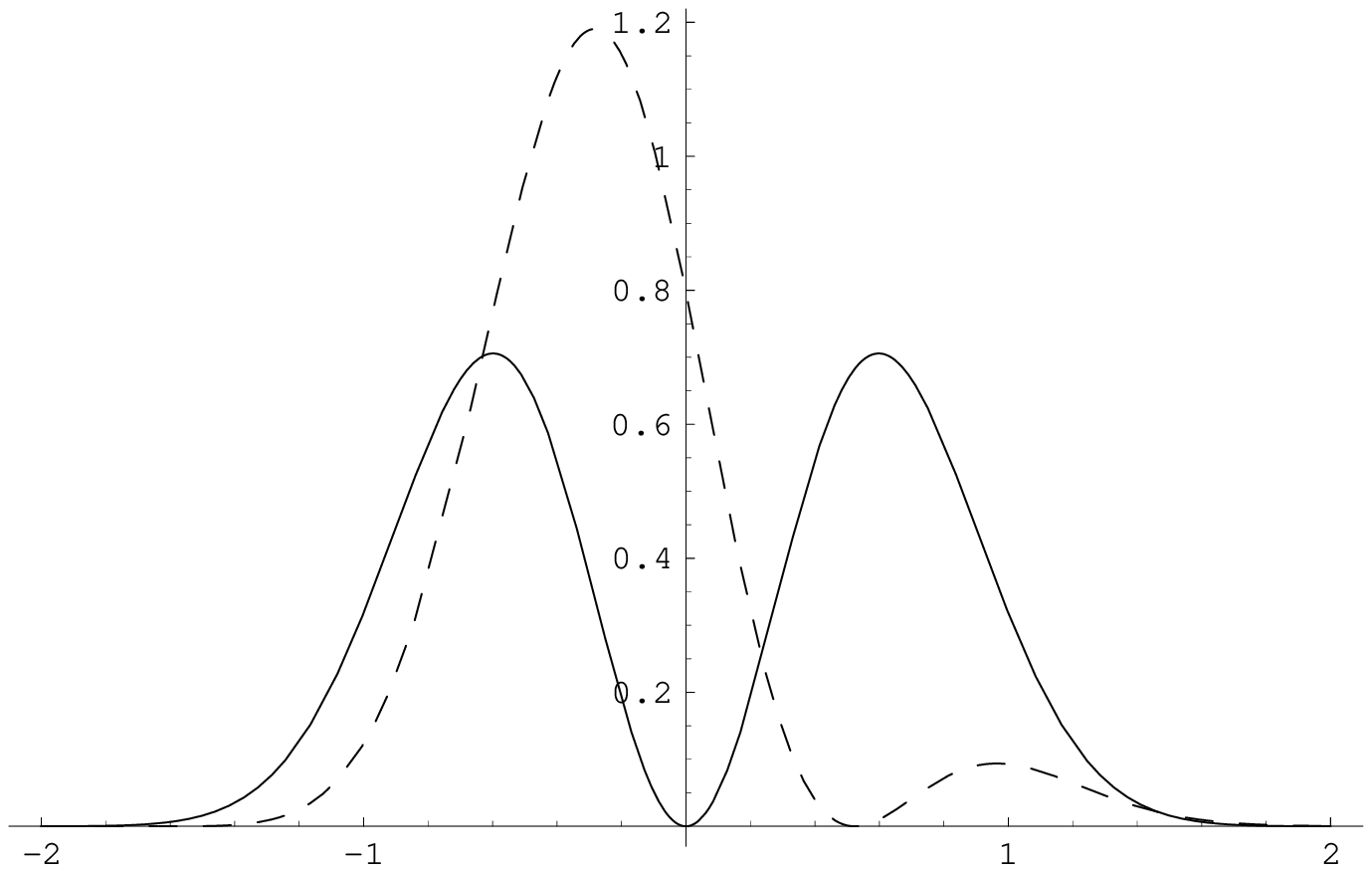}
\end{center}
\caption{The probability density for the Fourier transforms  of
the tactics $\mathcal{D}_{3}(-1)$ (the continuous line) i
$\mathcal{D}_{3}(\text{i})$ (the dashed line).} \label{intsie}
\end{figure}

The strategy $|\psi\rangle$, represented by the function $\langle
q|\psi \rangle$ or its transform $\langle p|\psi \rangle$,
expresses therefore the full relation of the player with the
market. The independent treatment of supply and demand aspects of
behaviours appearing on the market is not in  agreement with the
quantum description. As far as the interference phenomenon is
discussed, the tactics $\mathcal{D}_{\Delta}(z)$ do not commute
with the Fourier transformation hence it is not indifferent in
which representation, the supply or demand one, the player
performs the superposition of the strategies. The supply
probability density for the strategies which are the demand
superposition of the Gaussian strategies for four different values
of their relative phase, that is of the angle $\arg(z)$, are
illustrated in Figures~\ref{intosi} and \ref{intsie}.

In the case of a player acting as a commercial agent, the
derivative of the cumulative probability could be interpreted as
the intensity of the sale signal which is sent by the player to
the market when he is going to sell a goods purchased
beforehand\footnote{Market games differ a little depending on
whether the player announces the will of buying or accepts the
price of goods.}.

We notice that manipulating with the phase allows for a wide
choice of the trader's behaviour  which characterizes the supply
aspect of the tactics $\mathcal{D}_{\Delta}(z)$.

The tactics with $\arg(z)\negthinspace=\negthinspace0$ has the
clearly localized sale signal (Figure~\ref{intosi}), in the
opposition to two equivalent maxima of the sale tactic with
$\arg(z)\negthinspace=\negthinspace\pi$. For
$\arg(z)\negthinspace=\negthinspace\frac{3}{2}\pi$
(Figure~\ref{intosi}) we have that with the falling of the
price\footnote{The variable $p$ has been chosen so that it
diminishes with the increase of the unit price of  goods.}, the
buy signal  appears at first cautiously, but after a farther
decline, it reappears for the second time as a stronger one.

The opposite situation occurs for
$\arg(z)\negthinspace=\negthinspace\frac{1}{2}\pi$
(Figure~\ref{intsie}). Let us notice that just only the
measurement of the random variable $\mathfrak{p}$ allows to
distinguish between the strategy $\mathcal{D}_{3}(\text{i})$ and
$\mathcal{D}_{3}(-\text{i})$.

\section{The conditional demand}

\begin{figure}[h]
\begin{center}
\includegraphics[height=8.25cm, width=12cm]{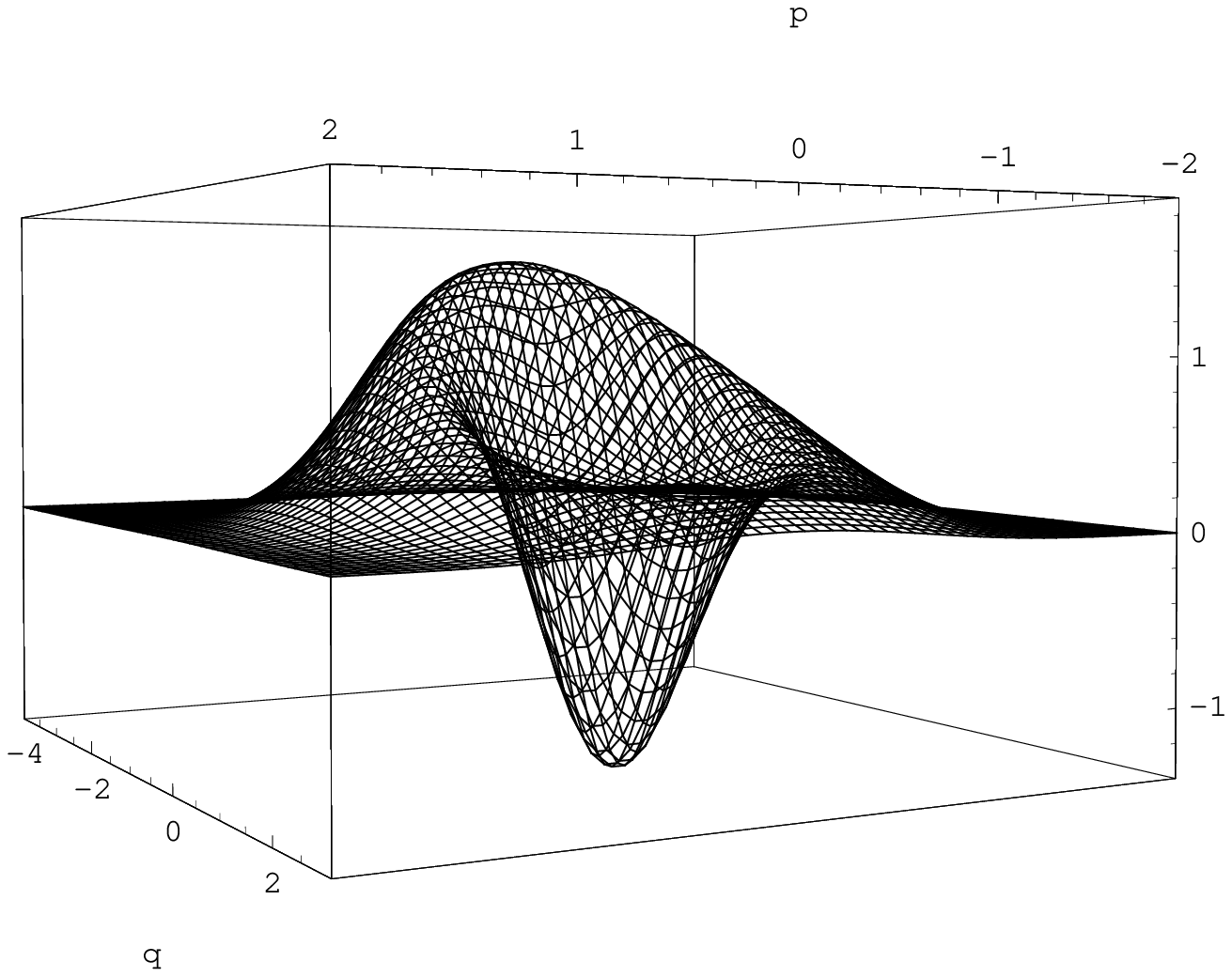}
\end{center}
\caption{Plot of the Wigner function for the tactic
$\mathcal{D}_{0.2}(-\sqrt{0.9})$.} \label{interpiec}
\end{figure}

An interesting thing is the examination of variety of aspects of
the strategies which is using the presentation of them in the
domain of the phase space $\{p,q\}$\index{The phase space.} with
the help of the Wigner function. For the pure state $|\psi\rangle$
this pseudo-probability density function can be calculated with
the following formula

\begin{eqnarray}
W(p,q)&:=& h^{-1}_E\int_{-\infty}^{\infty}e^{i\hslash_E^{-1}p x}
\;\frac{\langle
q+\frac{x}{2}|\psi\rangle\langle\psi|q-\frac{x}{2}\rangle}
{\langle\psi|\psi\rangle}\; dx\\
&=& h^{-2}_E\int_{-\infty}^{\infty}e^{i\hslash_E^{-1}q x}\;
\frac{\langle
p+\frac{x}{2}|\psi\rangle\langle\psi|p-\frac{x}{2}\rangle}
{\langle\psi|\psi\rangle}\; dx,
\end{eqnarray}
where $\frac{h_E}{2\pi}$ is the dimensionless economics
counterpart of the Planck constant.  Figure~\ref{interpiec}
presents the behaviour of the player who is operating with the
market tactic $\mathcal{D}_{0.2}(-\sqrt{0.9})$.

\begin{figure}[h]
\begin{center}
\includegraphics[height=6.25cm, width=9cm]{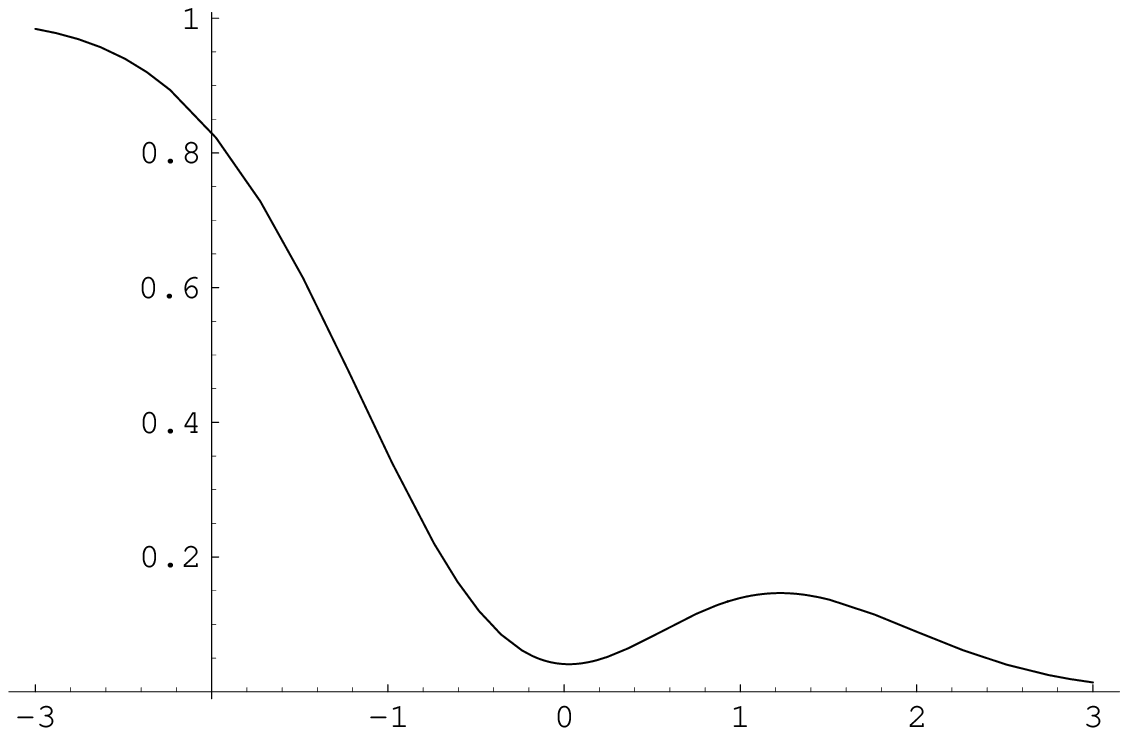}
\end{center}
\caption{The non-monotonous  conditional demand for the tactic
$\mathcal{D}_{0.2}(-\sqrt{0.9})$ (The integral curve for the
intersection of the surface from Figure~\ref{interpiec} with the
plain $p=0.4[\frac{\hslash_E}{\sigma}]$).} \label{interszes}
\end{figure}

Note that the asymmetric crater-like hollow in the diagram has the
minimum bellow zero, the fact which qualitatively distinguishes
the Wigner function from the supply and demand distributions for
models formulated in the realm of the classical probability theory
in which the measure of the probability has to be nonnegative. The
intersection of the surface of the diagram with the surface given
by $p\negthinspace=\negthinspace\text{constant}$ depicts the
conditional probability density which is the measure of the
probability for the withdrawal price of the player in the
situations when this price is constant during the act of selling.
The cross sections for the negative values of the Wigner function
are characteristic for the situation of the giffen-strategy. The
suitable integrals for these curves represent fully rational
situations for which the demand (or supply) cease to be a
monotonous  function. The example of such a reaction of the player
(it might be the rest of the world) is illustrated in
Figure~\ref{interszes}\index{The rest of the world @{\em reszta
świata}\/ (RŚ)}.
We observe here the lack of the property of the monotonicity for
the demand (or supply) curves (Giffen paradox).
In this context it is worthy to raise the question whether the
legendary captain Giffen, after observing a market anomaly which
is contradictory to the law of demand, has recorded the surprising
(although having an explanation) demand that decreases after the
fall of the price, or simply noticed
the destructive interference which had been the effect of a
careful demand transformation characteristic for a intelligent
(hence acting rationally) but poor consumer. The authors incline
towards the second answer. It has the superiority because of its
ability to be falsified \cite{13} which is a consequence of the
 precision   qualitative predictions for this
phenomenon made by the quantum theory. The case of the
non-monotonous conditional demand for interfering Gaussian
strategies has been characterized before. We recall that this
characteristic is the proper one for all non-Gaussian strategies.
Therefore it seems important to look after the conditions of the
market game under which the strategies described by the normal
distributions do not lead to the maximalization of value of the
intensity of the gain. They might explain the circumstances in
which we met the giffen-strategies.

\begin{figure}[h]
\begin{center}
\includegraphics[height=6.25cm, width=9cm]{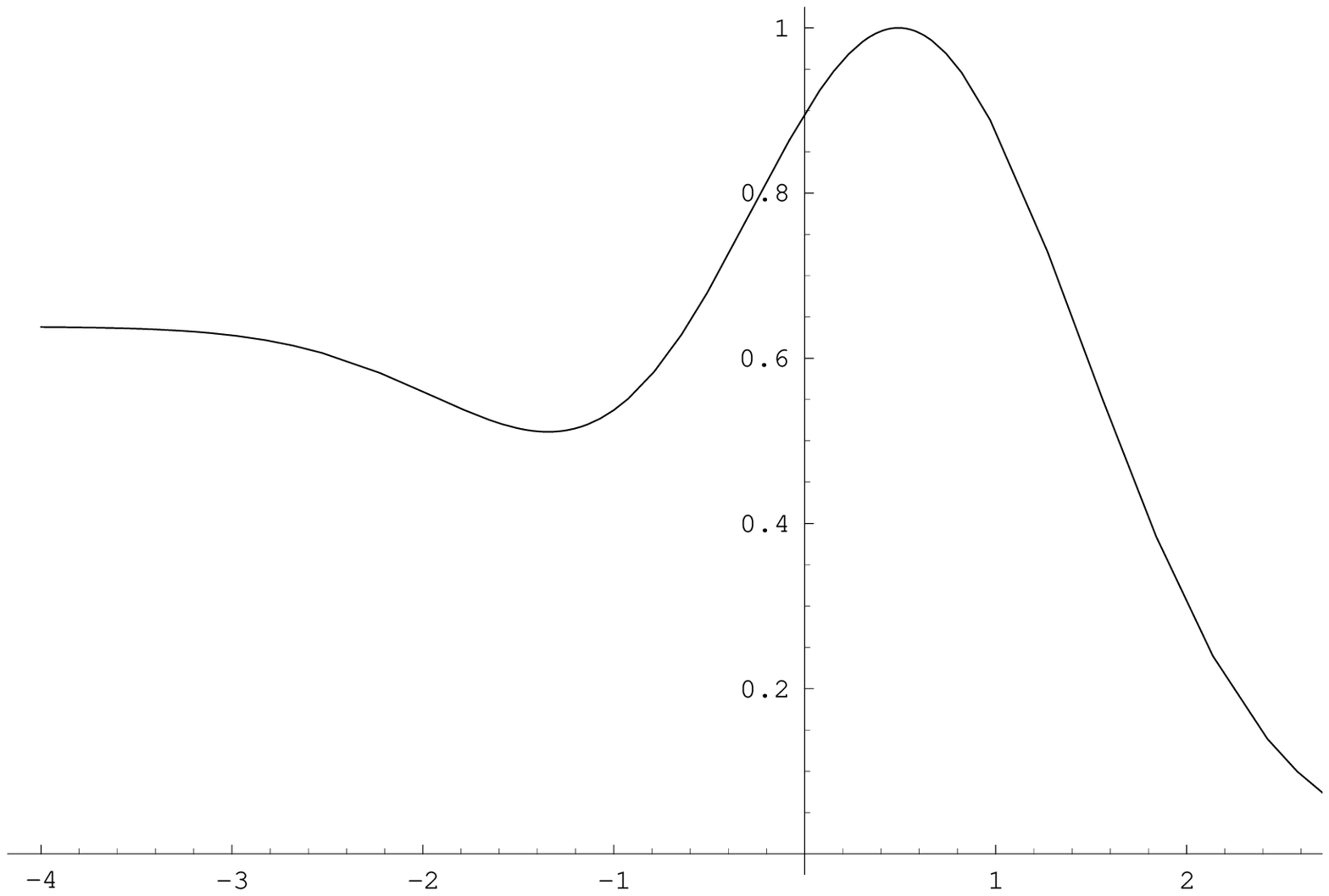}
\end{center}
\caption{The non-monotonous conditional demand with the maximum
for the nonzero value of the price. The tactic
$\mathcal{D}_{0.2}(-0.9^{-\frac{1}{2}})$ (the integral curve for
the cross section of the surface from Figure~\ref{interpiec} with
the plane $p=-0.2[\frac{\hslash_E}{\sigma}]$).} \label{hrysost}
\end{figure}

The careful reader has certainly noticed  other, besides the
non-monotonicity, characteristic of the integral curves for some
cross sections of the diagrams of the Wigner function. They have
 global maxima for definite values of the parameter
$q=q_{\max}$ (eg, for the surface in Figure~\ref{interpiec} with
the opposite orientation of the axis $p$). It means that there
exists a nonzero price of the commodity
$\text{e}^\frac{q_{\max}-q_0}{\sigma}$ under which the player
irreversibly losses his  interest in its buying. For example in
the capricious world of the fashion business such a kind of market
behaviours are frequently noticed; they are also known by the
owners of the rubbish auction\index{The rubbish auction}. The
example of a conditional demand curve is presented in
Figure~\ref{hrysost}. When, under the condition of  unchanging
tactic  player's estimation of the usefulness of the goods has
diminished properly, and afterwards the value of the parameter $p$
has been changed, then the new cross section shall be free from
this anomaly.

Both of the just discussed phenomena of the departures from the
law of demand and supply (and also others paradoxical properties)
disappear if only the cross sections are integrated over $p$. The
thus obtained  joint
cumulative probability of demand, being the integral of the square
of the module of the function $\langle q|\psi \rangle$, fulfills
the orthodox form of the law.

The above considerations have proved that the superposition of the
elementary quantum strategies has peculiarities which are well
known from the observations of  markets. The modeling of these
behaviours in the realm of the traditional probability theory
would be impossible for the same reason that made the physicists
to reject the classical picture of the world. In this context it
seems obvious that eg the designers of the software, which
automatically invests on  capital markets, sooner or later will be
confronted with the need of using  models of quantum market games.
Hence it is worthy to make a few comments on the unique efficacy
which is characteristic for the quantum tactics.

\section{Self-consistent nonlinear tactics}

In Ref\mbox{.} \cite{12} the nonlinear tactics
$\mathcal{D}_{E(\mathfrak{q})}$ operating on the Dirac strategy
which could be perceived as a limit of  Gaussian strategies for
$\sigma\negthinspace\rightarrow\negthinspace0$ were analyzed. Due
to the unique property which is connected with the fixed point
theorem \cite{12} it appears to be the method of the maximization
of the gain in the market game, in which the rest of the
world\index{The rest of the world@{\em The rest of the world}\/
(RŚ)}, being in a polarization \index{Polarization} compelling to
the exhibition of the offers of prices, accepts the passive tactic
$\mathcal{I}$ on its own Gaussian tactic. The tactic
$\mathcal{D}_{E(\mathfrak{q})}$ requires the measurement of the
mean $E(\mathfrak{q})$ hence it can not be used in a single game.
Yet, in the sequence of identical games, when a partner behaves in
a passive manner, we are able, in view of the past experiences, to
approximate the parameter $E(\mathfrak{q})$ so that in the
following one to shift our withdrawal price to this value. The
tactic $\mathcal{D}_{E(\mathfrak{q})}$ understood in this way is
then the unitary one, and being the quantum operation guarantees
to us the conditions required for being confidential. The just
mentioned fixed point theorem  assumes that  measure of the
probability given by the opponent's tactic is positive definite.
It is possible that if an obligatory passive opponent (e.g.~the
rest of the world is not able to react to the inconvenient
behaviour of only one small participant of the market) chooses the
giffen-strategy then the tactic $\mathcal{D}_{E(\mathfrak{q})}$
losses the quality of the most beneficial one. The infancy of
quantum computers do not hinder us from simulating the scenarios
of  quantum games on the traditional computers. In such situations
the use of the simplest quantum tactics which are similar to the
ones presented here should bring the completely new qualitative
effects. These tactics are more rich than the hardly similar to
the alive human market behaviour present algorithms for automatic
investment. Even today the quantum strategies might be used on
capital markets if only funds are raised to be managed in the
quantum manner. Also nothing is opposing to the possibility that
new markets appear, on which all parties of transactions might be
using quantum tactics. The physical quantum models offer the
solutions which are extremely stable in comparison with their
classical counterparts \cite{14}. So, perhaps the quantum markets
would protect the future society from negative consequences of the
deep oscillations of the prices which are characteristic for
nowadays economy.

For the needs of the time, when the principles of the functioning
of the market allow to analyze behaviors of partners of
transactions, the game, which is played with an opponent whose
tactic is reacting to our changes of the strategy, should be
examined. Such games posses the thermodynamic analogies and allows
to describe processes of reaching the market equilibrium. Because
of the extension of the subject, the analysis of the competitive
market tactics exceeds the limits of this paper which sketches
only the theme of the quantum market games.

\section{Conclusions}
When the player has at his disposal two pure strategies
$|\psi_0\rangle$ and $|\psi_1\rangle$ then the set of the
classical strategies in which (going from $|\psi_0\rangle$ to
$|\psi_1\rangle$) he can make changes in a continuous way, is
limited to the one dimensional set of the mixed strategies spanned
over the points $|\psi_0\rangle$ and $|\psi_1\rangle$. In
comparison to this, the quantum theory offers him extremely reach
possibilities. He or she faces a choice of a one from many
optional continuous curves for many of the paths connecting the
poles of the sphere $S_2$. Therefore such a "quantum player" puts
the qualitatively new challenges to an opponent. The opponent who
has at his disposal the knowledge based on the classical game
theory only, often is doomed from the start.\index{The market
tactics|)}
\def\urla{\href{http://econwpa.wustl.edu:8089/eps/get/papers/9904/9904004.html}{http://econwpa.wustl.edu:8089/eps/get/papers/9904/9904004.html}}
\def\urlb{\href{http://www.spbo.unibo.it/gopher/DSEC/370.pdf}{http://www.spbo.unibo.it/gopher/DSEC/370.pdf}}
\def\urlc{\href{http://www.comdig.org}{http://www.comdig.org}}

\end{document}